\documentclass{ws-p8-50x6-00}
\begin{document}

\title{Gauge Symmetry and Neural Networks}

\author{Tetsuo Matsui}

\address{Department of Physics, Kinki University, 
Higashi-Osaka, 8502 Japan
\\E-mail: matsui@phys.kindai.ac.jp}


\maketitle

\abstracts{
We propose a new model of neural network. It consists of spin
variables to describe the state of neurons as in the Hopfield model
and new gauge variables to describe the state of synapses.
The model possesses local gauge symmetry and resembles
lattice gauge theory of high-energy physics.
Time dependence of synapses describes the process of learning.
The mean field theory predicts a new phase corresponding to
confinement phase, in which  brain loses
ablility of learning and memory. }

\section{Introduction}
The Hopfield model of neural 
network\index{neural network} \cite{Hopfield}
succeeds to explain some basic 
functions of human brain such as associated 
memory\index{associated memory}. 
However, to be a more realistic model, at  least  the following points
should be taken into account;\\

\noindent
(1) Effects of external stimulations through eyes 
and ears on neurons.\\
\noindent
(2) Effects of time variations of synapses on neurons.\\

\noindent
The point (2) is essential to describe the function of 
learning\index{learning},
since  the   possible patterns to memorize are completely
determined according to the strengths of synapse connections 
among neurons as  long as they are time independent.
Their time dependence induces the process of learning itself.

In Sect.2, we review the Hopfield model briefly. 
In Sect.3, we propose a new model of neural network, in which the 
strengths of synapse connections are regarded as gauge connections
and vary in time according to the gauge principle. 
By using the mean field theory, we see that the  model predicts
a new state of brain in which both learning and memory are impossible. 
In Sect.4, we present future outlook.

\section{Hopfield model\index{Hopfield model}}

Let us review the framework of the Hopfield model briefly.
Its energy $E_H(\{S_i\})$ is given by
\begin{eqnarray}
E_H(\{S_i\}) & = & -\frac{1}{2}\sum_{i = 1}^{N}
\sum_{j = 1}^{N}J_{ij}S_i S_j,
\label{EH}
\end{eqnarray}
where $S_i = \pm 1$ is the Ising spin variable to describe
the state of $i$-th neural cell ($i = 1,2, \cdots, N$) as 
 $S_i = 1$; excited,\ $S_i = -1$; unexcited. $J_{ij}$ is a
 given real constant that expresses
the strength of synapse connection for the signal 
propagating from the 
$j$-th cell to the $i$-th cell. 

The time evolution of  $S_i(t)$ for every discrete time
interval $\epsilon$ (often set unity) is governed by 
the following equation;
\begin{eqnarray}
S_i(t+ \epsilon) & = & sgn[-\frac{\partial E_H}
{\partial S_i}(t)] 
= sgn[\sum_j J_{ij}S_j(t) ].
\label{dynamical}
\end{eqnarray}
Thus,
$J_{ij} > 0 $ tries to proliferate (un)excited cells, while
$J_{ij} < 0 $ prefers mixtures of excited and unexcited ones. 
If the system converges into certain configuration of $\{S_i\}$
after a sufficiently long time, it corresponds to recalling
certain pattern.
Such a configuration should be a stationary point of 
$E_H$, i.e., $\partial E_H/\partial S_i = 0$ for every $i$.
All these configurations are determined once $J_{ij}$ are given. 

Practically speaking, the rule (\ref{dynamical}) may not 
necesssarily
hold all the time interval due to anavoidable error in signal 
propagations. Such a situation may be simulated by adding
random noises $\eta_i(t)$ into the square bracket in 
the right-hand side of 
(\ref{dynamical}), whose strength can be identified as a 
fictitious "tempeteture" $T$. If $T$ is large, the error in
signal propagations occurs frequently. Thus it is interesting to
study statistical mechanics of the system $E_H$ by using
Boltzmann distribution. The partition function $Z_H$ is given by  
\begin{eqnarray}
Z_H & = & \prod_i \sum_{S_i =\pm 1} \exp(-\beta E_H), \ \ \beta 
\equiv 1/T. 
\end{eqnarray}
In case that all $J_{ij}$ are positive, the system has two phases;\\

\noindent
-  Ferrofagnetic phase below certain critical temperature
$T_c$, $T < T_c$,
 in which there is a long-range order and the average 
$\langle S_i \rangle \neq 0$.\\

\noindent
- Paramagnetic phase  above $T_c$, $T > T_c$, in
which $S_i$ are random and $\langle S_i \rangle = 0$.\\

\noindent
The ferromagnetic phase corresponds to the state of clear
memory, while in the paramagnetic  phase
 no definite patterns can persist.
If $J_{ij}$ is complicated, there arises 
a spin-glass phase\index{spin glass} as we shall see.

Explicitly, let us fix $J_{ij}$ according to the Hebb's rule as 
\begin{eqnarray}
J_{ij}  & =&   \frac{1}{N}\sum_{\alpha = 1}^M 
\xi_{i}^{\alpha} \xi_{j}^{\alpha},
\end{eqnarray}
where we prepare $M$ patterns $S_i = \xi_i^{\alpha}\ 
(\alpha = 1,\cdots M)$ to recall. The replica method 
gives rise to the phase diagram shown in Fig.1.\cite{Amit} 
Each phase  is explained in Table 1.

\begin{figure}[t]
\begin{minipage}{.2\linewidth}
\epsfxsize=4pc 
\epsfbox{null.eps}
\end{minipage}
\begin{minipage}{.7\linewidth}
\epsfxsize=15pc 
\epsfbox{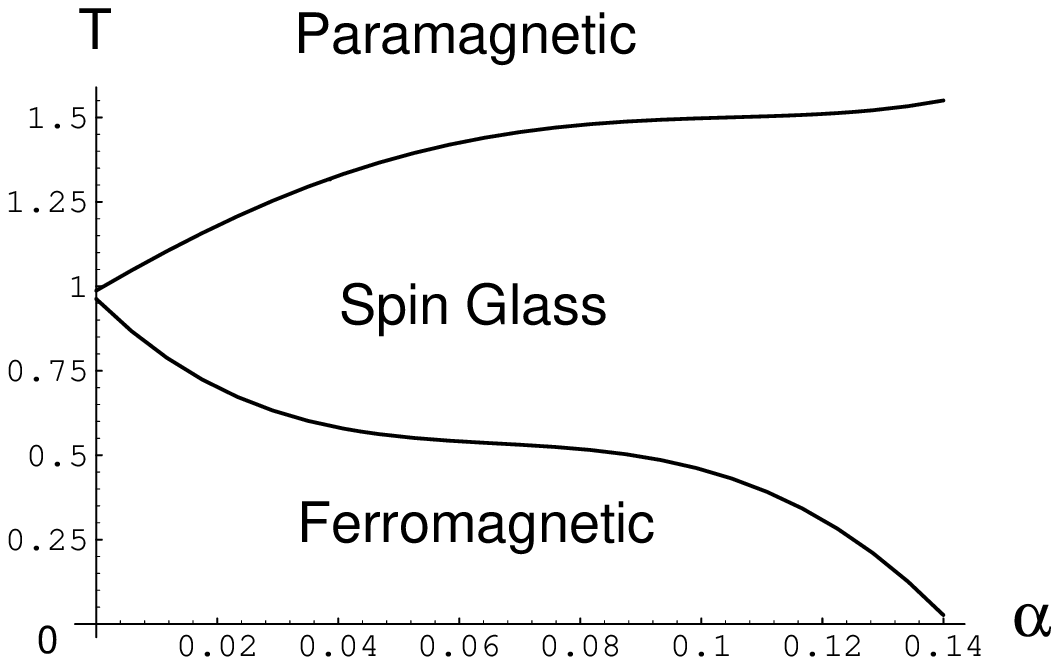} 
\end{minipage}
\caption{Phase structure of the Hopfield model in 
$\alpha(\equiv M/N)-T$ plane.
\label{fig1}}
\end{figure}

\hspace{0.5cm}
\begin{center}
Table 1. Phases of the Hopfield model. \\
\begin{tabular}{|c|c|c|c|}
\hline
    Phase  & $\sum_i \langle S_i \rangle $  
    & $\sum_i \langle S_i \rangle^2 $ &  Property    \\
\hline
Ferromagnetic& $\neq 0$ & $\neq 0$ & memory     \\
\hline
Spin glass & $0$ & $\neq 0$ & false memory    \\
\hline
Paramagnetic& $0$ & $0$ &  no memory   \\
\hline
\end{tabular}
\end{center}

\section{New model with local gauge symmetry\index{gauge symmetry} }

\begin{figure}[t]
\begin{minipage}{.2\linewidth}
\epsfxsize=4pc 
\epsfbox{null.eps}
\end{minipage}
\begin{minipage}{.7\linewidth}
\epsfxsize=18pc 
\epsfbox{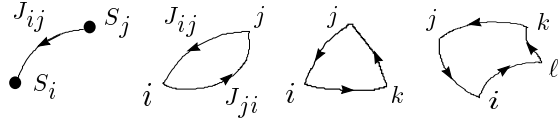} 
\end{minipage}
\caption{Graphical representation of each term in $E$ of (\ref{e}).
\label{fig2}}
\end{figure}

As pointed out in Sect.1, to incorporate the function of learning,
one needs the time variation of $J_{ij}$. There are various 
approaches for this point. 
Below, we regard both $S_i$ and 
$J_{ij}$ as dynamical variables and treat them on an equal footing.
Let us assume that their 
time dependence is controlled so as to reach a local minimum 
of the new energy $E(\{S_{i},J_{ij}\})$. To determine $E$, 
we impose the conidtion that $E$ is local
gauge invariant under the following gauge transformation; 
\begin{eqnarray}
&&S_i  \rightarrow S'_i \equiv V_i S_i, \ \ \   J_{ij}  
\rightarrow J'_{ij} \equiv V_i J_{ij} V_j, \ \ \  
E(\{S'_i, J'_{ij}\}) =  E(\{S_i, J_{ij}\}),
\end{eqnarray}    
where $V_i = \pm 1$ is the Z(2) variable associated with
$i$-th cell.
Since $J_{ij}$ describes the state of the synapse connecting
$i$-th and $j$-th cells, it is natural to regard it as
the  connection of  gauge theory.  The neural network
may possesses certain conservative 
quantity in association with the long-term memory.
The local gauge symmetry we address may respect such a 
conservation law. This point will be reported in detail in a 
separate publication.\cite{Sakakibara}
It is often stressed that the connections $J_{ij}$ and $J_{ji}$
are independent (asymmetric). 
Then  a general form of $E(\{S_{i},J_{ij}\})$ and 
the partition function $Z$ may be given by
\begin{eqnarray}
E &=& -\frac{1}{2}\sum_{i,j}S_i J_{ij} S_j 
+ \frac{g_2}{2} \sum_{i,j}J_{ij}J_{ij} \nonumber\\
&+& \frac{g_3}{3!} \sum_{i,j,k}J_{ij}J_{jk}J_{ki}
+ \frac{g_4}{4!} \sum_{i,j,k,\ell}J_{ij}J_{jk}
J_{k\ell}J_{\ell i} + \cdots
\nonumber\\
Z & = & \prod_i\sum_{S_i = \pm 1} \prod_{i\neq  j}
 \int dJ_{ij} \exp (-\beta E), \ \ \beta \equiv 1/T.
\label{e}
\end{eqnarray}
Each term of $E$ is gauge invariant since $V_i^2 = 1$,
and depicted in Fig.\ref{fig2}.
 
$E$ takes a form very similar to the lattice gauge 
theory\index{lattice gauge theory}
in particle physics, where $S_i$ corresponds to  a
matter field and $J_{ij}$ to an exponentiated gauge field.
 If the parameters $g_2, g_3, g_3,...$
are set zero, $E$ reduces to $E_H$ of (\ref{EH}).

\subsection{Model I}

\begin{figure}[t]
\begin{minipage}{.3\linewidth}
\epsfxsize=4pc 
\epsfbox{null.eps}
\end{minipage}
\begin{minipage}{.7\linewidth}
\epsfxsize=12pc 
\epsfbox{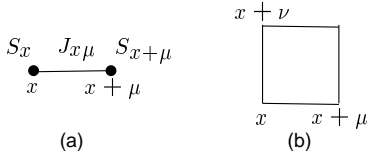} 
\end{minipage}
\caption{
Graphical representation of $E$ of (\ref{e1}). 
(a) $\lambda$ term. (b) $g^{-2}$ term.
 \label{fig3}}
\end{figure}

To be explicit, we need to specify the model further.
Let us first consider the 
$Z(2)$ Higgs gauge model\index{Z(2) Higgs gauge model}
on a 3D cubic lattice,
\begin{eqnarray}
E_{\rm I} &=& -\lambda \sum_{x}\sum_{\mu = 1}^3 
S_{x+\mu}J_{x\mu} S_x - \frac{1}{g^2} 
\sum_{x} \sum_{\mu < \nu} J_{x\mu}J_{x+\mu,\nu}
J_{x+\nu,\mu}J_{x\nu},
\nonumber\\
Z_{\rm I} & = & \prod_x\sum_{S_x = \pm 1} \prod_{x,\mu} 
\sum_{J_{x\mu}= \pm 1} \exp (-\beta E_{\rm I}) \equiv 
\exp(-\beta F_{\rm I}),  
\label{e1}
\end{eqnarray}
where $x$ denotes 
the lattice site on which $S_x$ lives, and 
$\mu\ (=1,2,3)$ denotes both the direction and 
the unit vector. We consider only  the connections between
the nearest-neighbor sites $(x,x+\mu)$ and treat them as
symmetric $Z(2)$ variable on a link $(x,x+\mu)$;
$J_{x,x+\mu} = J_{x+\mu,x} \equiv J_{x\mu} = \pm 1$.
The $\lambda$ term and the $1/g^2$-term is depicted 
in Fig.\ref{fig3}.

The time evolution of $J_{x\mu}$ may be given by the 
similar rule as (\ref{dynamical}), 
\begin{eqnarray}
S_x(t+ \epsilon) & = & sgn[- 
\frac{\partial E_{\rm I}}{\partial S_x}(t) + \eta_x(t)], \nonumber\\ 
J_{x\mu}(t+ \alpha \epsilon) & = & sgn[-
\frac{\partial E_{\rm I}}{\partial J_{x\mu}}(t) + \zeta_{x\mu}(t)],
\label{dynamicalI}
\end{eqnarray}
where $\alpha$ sets the ratio of the two time scales for
$S_x$ and $J_{x\mu}$.
We report our study of (\ref{dynamicalI}) elsewhere.\cite{Kemuriyama}

\begin{figure}[t]
\begin{minipage}{.25\linewidth}
\epsfxsize=4.5pc 
\epsfbox{null.eps}
\end{minipage}
\begin{minipage}{.7\linewidth}
\epsfxsize=16pc 
\epsfbox{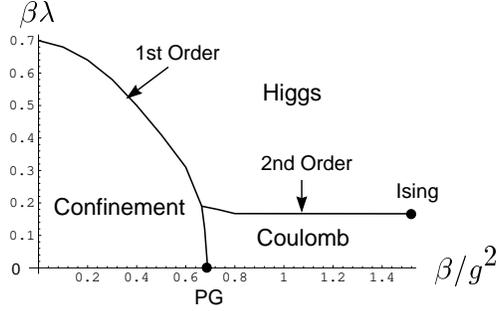} 
\end{minipage}
\caption{
Phase diagram of Model I of (\ref{e1}). The point marked as
Ising locates the second-order transition point of the 
Ising model\index{Ising model}.
The point PG locates the first-order transition point of the Z(2)
pure gauge theory. 
\label{fig4}}
\vspace{-0.5cm}
\end{figure}

Below we study the phase diagram of $E_{\rm I}$ by using the 
mean field theory, which is formulated as a varational
principle as follows. Let us  introduce a variational
energy $E_0$. Then the Jensen-Peierls
inequality gives rise to 
\begin{eqnarray}
F_{\rm I}  & \stackrel{\textstyle <}{\_} &  F_0 + 
\langle E_{\rm I} - E_0 \rangle_0,
\nonumber\\
Z_0 & = & {\rm Tr}\ \exp (-\beta E_0) \equiv 
\exp(-\beta F_0), \nonumber\\
\langle O \rangle_0 &=& Z_0^{-1}\ {\rm Tr}\ O\ 
\exp (-\beta E_0),
\end{eqnarray}
where Tr implies $\prod_x\sum_{S_x = \pm 1} \prod_{x,\mu} 
\sum_{J_{x\mu}= \pm 1}$.
We  choose the variational parameters in $E_0$ so that
the right-hand-side of inequality reaches the  minimum.
For $E_0$  we assume the translational invariance of mean fields
and employ the single-site and single-link energy, 
\begin{eqnarray}
E_0 &=& -\sum_{x}\sum_{\mu} W_{x\mu} J_{x\mu} -\sum_x h_x S_x,
\end{eqnarray}
with the two variational parameters, $W_{x\mu}=W$ and  $h_x = h$.
The result is given in  Table 2 and Fig.\ref{fig4}.

\hspace{0.5cm}
\begin{center}
Table 2. Phases of Model I of (\ref{e1}). \\
\begin{tabular}{|c|c|c|c|c|c|}
\hline
Phase & $\langle S_{x} \rangle $ &
$\langle J_{x\mu} \rangle $ & Memory & Learning  & Hopfield Model 
\\  
\hline
 Higgs & $\neq 0$ & $\neq 0$ & yes & yes & Ferromagnetic
\\ 
\hline
 Coulomb & $0$ & $\neq  0$ &no &  yes & Paramagnetic
\\ 
\hline
 Confinement & $0$ & $ 0$ & no & no & not available
\\ 
\hline
\end{tabular} \\
\end{center}

\newpage

As shown in the Table 2, one may take $\langle S_{x} \rangle $
as an order parameter to judge whether the system succeeds to
recall definite patterns, and $\langle J_{x\mu} \rangle $
to judge whether the system is able to learn some new patterns.
 In the confinement phase, neither memory nor learning is possible.
This phase is missing in the Hopfield  model. 
 
We note that  Monte Carlo simulations of the 3D Z(2) Higgs 
gauge model exhibit these 
three phases, but the phase boundary of Higgs and confinement 
phases does not continue to $\beta/g^2 = 0$ but terminates
at some finite value. These two phase can be reached
each other smoothly by contouring the end point.
This "complementarity" reflects that $|J_{x\mu}| = 1$
and is proved by rigorous
treament,\cite{Fradkin} but not predicted correctly in our 
variational treatment.\cite{Drouffe}
 
To what extent are these results trustworthy? 
To answer this question, 
we intoroduce and study other two models, Model II and Model III.

\subsection{Model II}

The framework and the energy $E_{\rm II}$ of Model II 
is same as Model I,
but we allows  additional state $J_{x\mu} = 0$, which describes
the possibility that the connection between 
$x$ and $x+\mu$ is missing;
\begin{eqnarray}
E_{\rm II} &=& E_{\rm I}, \ \ \ 
Z_{\rm II}  = 
 \prod_x\sum_{S_x = \pm 1} \prod_{x,\mu} 
\sum_{J_{x\mu}= 0, \pm 1} \exp (-\beta E_{\rm II}).
\label{e2}
\end{eqnarray}

\begin{figure}[t]
\begin{minipage}{.2\linewidth}
\epsfxsize=5pc 
\epsfbox{null.eps}
\end{minipage}
\begin{minipage}{.7\linewidth}
\epsfxsize=14pc 
\epsfbox{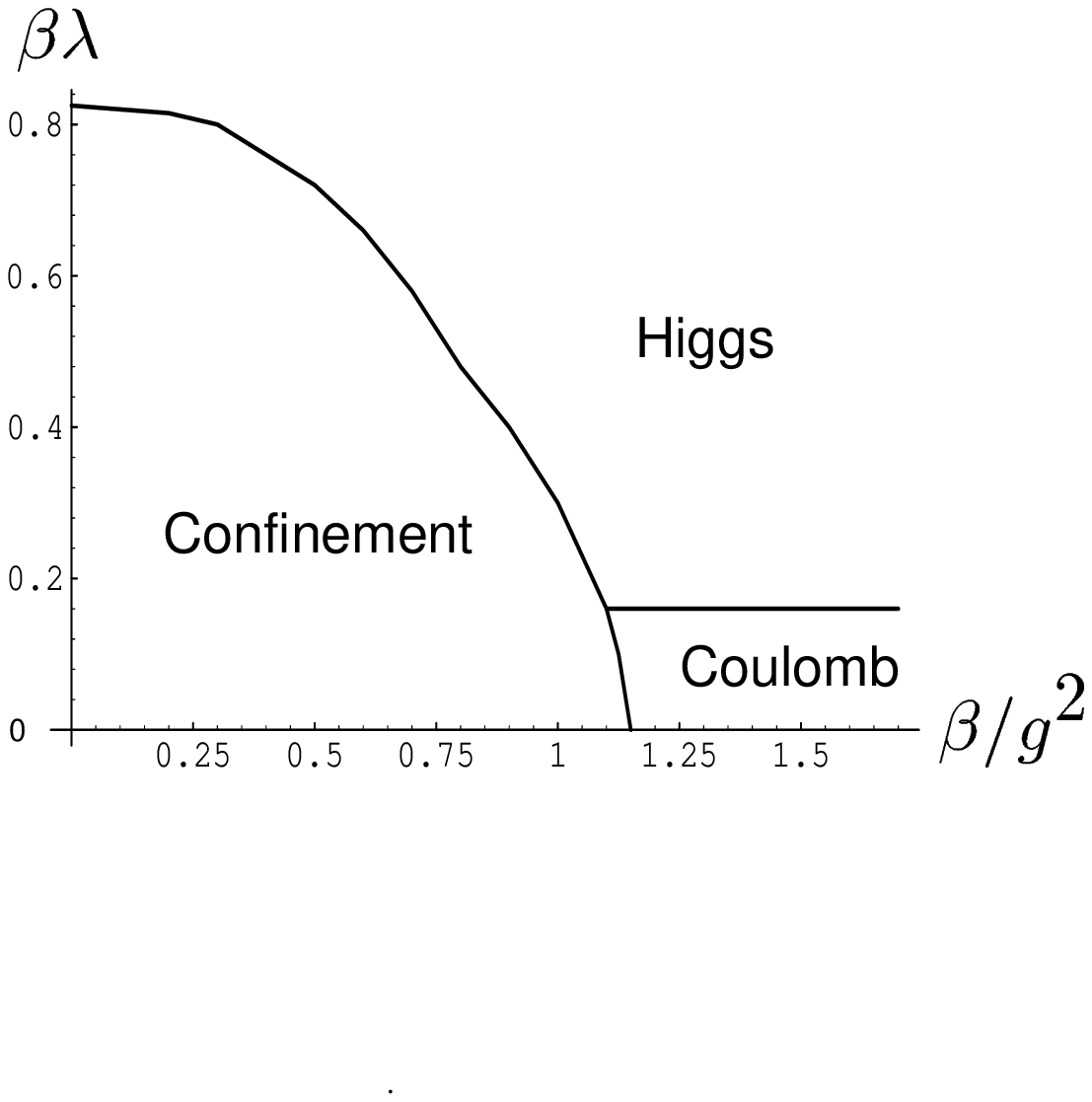} 
\end{minipage}
\vspace{-2cm}
\caption{
Phase diagrams of Model II of (\ref{e2}). 
\label{fig5}}
\end{figure}

The phase diagram  calculated by the similar mean field theory
is shown in Fig.\ref{fig5}.  
The global structure remains the same as Fig.\ref{fig4},
although the region of the confinement
phase is enlarged as expected since the added states clearly
favor this phase. 

\newpage

\subsection{Model III}

In Model III, 
 we intorduce two independent $Z(2)$ variables $J_{x\mu}$
 and $\bar{J}_{x\mu}$ for the  synapse between $x$ and $x+\mu$ 
 to take their independence into account as
\begin{eqnarray}
 J_{x\mu} \equiv J_{x, x+\mu}, \ \ 
 \bar{J}_{x\mu} \equiv J_{x+\mu,x}. 
\end{eqnarray}
We also define $J_{x,-\mu} \equiv J_{x-\mu,x}$.
The energy $E_{\rm III}$ is then given by
\begin{eqnarray}
E_{\rm III} &=& -\lambda \sum_{x}
\left( \sum_{\pm\mu } \bar{J}_{x,\pm\mu}S_{x\pm\mu}\right)
\left( \sum_{\pm\nu } J_{x,\pm\nu}S_{x\pm\nu}\right)
\ \ \ \ \ \ \ \ \ \ \ \ \ \ \ \ \ \ \ \ {}\nonumber\\
&- &\frac{1}{g^2} 
\sum_{x} \sum_{\mu < \nu}  \left[ \bar{J}_{x\mu}
\bar{J}_{x+\mu,\nu}J_{x+\nu,\mu}J_{x\nu}
+ (\mu \leftrightarrow \nu)\right].
\label{e3}
\end{eqnarray}
We note that the expression $J_{ij} S_i S_j$ in
$E_{\rm H, \ I, \ II}$ washes out the asymmetry $J_{ij}
\neq J_{ji}$, while the fist term of (\ref{e3}) reflects it.
 Each term in $E$ is depicted in Fig.6.

\begin{figure}[b]
\begin{minipage}{.25\linewidth}
\epsfxsize=4pc 
\epsfbox{null.eps}
\end{minipage}
\begin{minipage}{.7\linewidth}
\epsfxsize=16pc 
\epsfbox{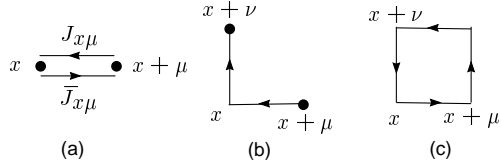} 
\end{minipage}
\caption{
Graphical representation of Model III of (\ref{e3}).
(a)  $J_{x\mu}$ and  $\bar{J}_{x\mu}$; (b) $\lambda$ term;
(c) $g^{-2}$ term.  
\label{fig6}}
\end{figure}

\begin{figure}[t]
\begin{minipage}{.2\linewidth}
\epsfxsize=4pc 
\epsfbox{null.eps}
\end{minipage}
\begin{minipage}{.7\linewidth}
\epsfxsize=15pc 
\epsfbox{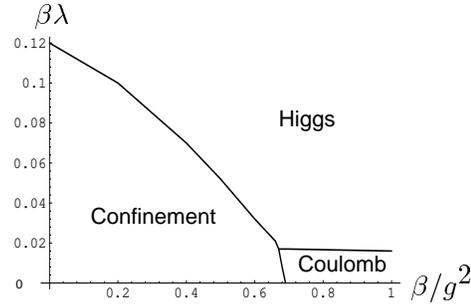} 
\end{minipage}
\caption{
Phase diagrams of Model III of (\ref{e3}). 
\label{fig7}}
\end{figure}

The phase diagram  in mean field theory
is shown in Fig.\ref{fig7}.  
The global structure still remains unchanged,
although the region of the confinement
phase is diminished considerably. This may be understood
since the first term in $E_{\rm III}$ is bilinear in $J_{ij}$
in contrast  with $E_{\rm H, \ I, \ II}$, and favors
nonvanishing $J_{ij}$.

\section{Summary and outlook}

Our results may be summerized with some future 
outlook as follows; \\

\noindent
- Due to the dynamical variables $J_{ij}$, a new (confinement) 
phase appears at high temperatures, which describes the new
state of no ability of learning and memory.\\

\noindent
- To decribe the spin-glass phase, further study of 
long-range correlation and/or frustrations is necessary.\\

\noindent
- Relaxing of $J_{ij} =\pm1(, 0) $ to $-\infty < J_{ij} < \infty$ 
may be interesting, but requires a detailed form of the energy.\\

\noindent
- Study of the time evolution of $J_{ij}$ and $S_i$ 
may describe the mechanism of learning such as the
process to forget the patterns. \\



\noindent
- Study of the effect of local gauge symmetry on brain 
function on a "quantum" level is interesting. Introduction of 
gauged versions of quantum brain 
dynamics\index{quantum brain dynamics} \cite{QBD} and 
cellular automata with Penrose's idea \cite{Penrose} may 
be the first step.

\newpage
\noindent
{\bf Acknowlegement}\\

I thank Prof. Hagen Kleinert for fruitful discussion on various 
fields of physics during  my pleasant stay at FU Berlin in 1983-1990.
I also appreciate discussions with Dr. Kazuhiko Sakakibara  
and Mr. Motohiro Kemuriyama.

\end{document}